\documentclass[notspecified,article,submit,pdftex,moreauthors]{Definitions/mdpi}
\providecommand{\tightlist}{\setlength{\itemsep}{0pt}\setlength{\parskip}{0pt}}

\firstpage{1}
\pubvolume{1}
\issuenum{1}
\articlenumber{0}
\pubyear{2026}
\copyrightyear{2026}
\datereceived{}
\daterevised{}
\dateaccepted{}
\datepublished{}

\definecolor{revref1}{rgb}{0.55,0.0,0.55}
\def\xillver{{\sc xillver}}
\def\relxill{{\sc relxill}}

\Title{Black-Hole Spin Measurements from X-ray Reflection Spectroscopy: Quality Criteria and Community Recommendations}

\Author{Javier A. Garc\'ia $^{1,2,*}$\orcidA{}, Riley Connors$^{3}$, Laura W. Brenneman$^{4}$, and James F. Steiner $^{5}$}
\AuthorNames{Javier A. Garc\'ia, Riley Connors, Laura W. Brenneman, and James F. Steiner}

\address{%
$^{1}$ \quad NASA Goddard Space Flight Center, Greenbelt, MD, USA; javier.a.garcia@nasa.gov\\
$^{2}$ \quad California Institute of Technology, Pasadena, CA, USA; javier@caltech.edu\\
$^{3}$ \quad Department of Physics, Villanova University, Villanova, PA, USA; riley.connors@villanova.edu\\
$^{4}$ \quad Center for Astrophysics | Harvard \& Smithsonian, Cambridge, MA, USA; lbrenneman@cfa.harvard.edu\\
$^{5}$ \quad Center for Astrophysics | Harvard \& Smithsonian, Cambridge, MA, USA; james.steiner@cfa.harvard.edu}
\corres{Correspondence: javier.a.garcia@nasa.gov}

\abstract{X-ray reflection spectroscopy provides one of the most effective
electromagnetic routes to measuring the dimensionless spin parameter of
accreting black holes. The method has produced spin constraints for both
stellar-mass black holes in X-ray binaries and supermassive black holes in
active galactic nuclei, and it is central to the science objectives of present
and future X-ray telescopes. At the same time, this method is vulnerable to a
set of coupled observational and modeling systematics: continuum-reflection
degeneracy, insufficient passband, unresolved distant reflection or absorption,
detector effects, source variability, accretion-state dependence, and
assumptions built into the reflection model itself. This article
synthesizes the talks and discussions held at the 2025 Wake Forest workshop
\emph{Recent Progress on Black Hole Spin Measurements Across the
Electromagnetic and Gravitational Spectra}; motivated by them, we propose a
practical quality-control framework for assessing whether a published
reflection-based spin measurement should be treated as robust, provisional, or
not assessable from the published information alone. We organize the problem
around three pillars: \emph{detectability}, meaning that the relativistic
reflection signal is unambiguously present in the data; \emph{uniqueness},
meaning that the relativistic component can be separated from the continuum,
from distant reflection, absorption, and instrumental effects; and
\emph{robustness}, meaning that the inferred spin is stable against plausible
changes in model assumptions, data selection, and accretion-state treatment. We
then translate these principles into applicable criteria, a tiered
quality-classification scheme, and a reporting checklist for future analyses.
The quantitative calibration of each criterion requires a dedicated
campaign of simulations over realistic scenarios, which we outline here and
defer to a companion publication. We aim to define a reproducible path toward
a curated, community-maintained compilation of reliable spin constraints and to
guide the implementation of reflection spectroscopy in the high-throughput,
high-resolution era.}

\usepackage{xspace}

\begin{document}

\section{Introduction}\label{introduction}

Black-hole spin is a fundamental parameter of the Kerr solution to Einstein's
general relativity. Its measurement provides insights into black hole growth,
accretion history, and angular-momentum exchange. In black holes of
stellar-mass, spin encodes the outcome of core collapse, binary evolution, and
corresponding accretion
\citep{KingKolb1999,FullerMa2019,Qin2018,Fragos2015,MillerMiller2015}. In
supermassive black holes, spin is shaped by mergers, coherent or chaotic
accretion. Spin can regulate feedback and, therefore, connects strong-gravity
physics to galaxy evolution
\cite{Volonteri2005,KingPringle2006,BertiVolonteri2008,Sesana2014}. Unlike
mass, in electromagnetic systems spin cannot be measured dynamically in a
straightforward way. However, it can be inferred from its effects on matter and
radiation in the strong-field region.

\begin{figure}[H]
\centering
\includegraphics[width=\textwidth]{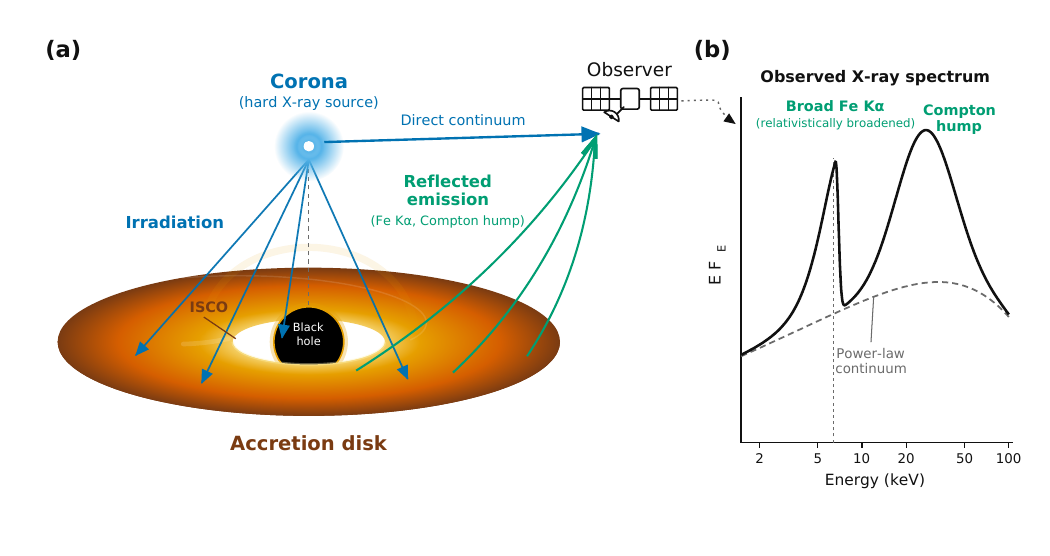}
\caption{Concept of X-ray reflection spectroscopy. Hard X-rays from a compact
corona illuminate an optically thick accretion disk, producing a reflection
spectrum with fluorescent lines (notably Fe K-shell transitions) and a Compton
hump due to electron scattering. Relativistic Doppler shifts, gravitational
redshift, and light bending imprint characteristic broadening that constrains
the inner disk radius (ISCO), enabling inference of the black hole spin
$a_*$.\label{fig:reflection}}
\end{figure}

{\em X-ray reflection spectroscopy} is one of the leading electromagnetic
methods to estimate black hole spin \citep[see][for recent
reviews]{Reynolds2021,KaraGarcia2025}. Hard X-ray photons from a compact corona
irradiate the optically thick accretion disk and produce a reflection spectrum
containing fluorescent lines, absorption edges, and Compton-scattered continuum
emission. The Fe K complex (i.e. inner-shell atomic transitions) and the
Compton hump (produced by electron scattering in the disk's atmosphere) are the
most prominent manifestations of reflection and are thus capable of serving as
powerful diagnostics (Fig.~\ref{fig:reflection}, left panel). If the reflecting
disk extends to the innermost stable circular orbit (ISCO), the degree of
gravitational redshift, Doppler broadening, and light bending imprinted on the
reflection spectrum can be used to infer the inner disk radius and hence the
spin (Fig.~\ref{fig:reflection}, right panel). In these models the spin is
represented by the dimensionless parameter $a_* = cJ/GM^2$, where $J$ is the
black hole's angular momentum, $c$ is the speed of light, and $G$ is the
universal gravitational constant. This particular parameterization implies that
the reflection method is independent of the black hole mass, making it
applicable to virtually any accreting system under the right conditions. This
picture underlies modern reflection models such as \xillver\ and \relxill\ and
has been extensively reviewed in the literature
\cite{Brenneman2006,Brenneman2013,Reynolds2014,Garcia2013ApJ768,Garcia2014ApJ782,Dauser2014,Dauser2016}.

The X-ray reflection method is powerful precisely because it is sensitive to
the innermost accretion flow; however, other physical phenomena which manifest
in the same energy range complicate its viability. A broad residual near Fe K
does not guarantee a reliable spin measurement; a Compton hump does not
automatically guarantee a constraint on relativistic reflection; and a formally
small statistical uncertainty on spin does not necessarily imply accuracy. The
posterior probability distribution of the spin derived from a Bayesian analysis
can be artificially narrow if the continuum is oversimplified, if distant
reflection or absorption is unresolved, if the passband does not also constrain
the continuum and Compton hump, if pile-up or calibration effects distort the
line profile, or if the adopted reflection model lacks relevant physics or is
incorrectly applied. These issues are not failures of the method; they are the
expected consequences of applying a highly constrained physical model to
complex astrophysical data.

The 2025 Wake Forest
workshop\footnote{\url{https://www.cardenas.sites.wfu.edu/spinworkshop/}} was
organized to bring together electromagnetic, gravitational-wave, and
horizon-scale-imaging communities working on black hole spin. The workshop
emphasized the need for self-consistent theoretical frameworks, the
quantification of selection effects, and the propagation of model uncertainties
into mission-level science requirements. For X-ray reflection, these themes
translate into a concrete need: the community requires an explicit and
reproducible way to determine if a spin measurement is sufficiently reliable
for population studies, mission planning, and comparisons with
gravitational-wave spin distributions. Here, we aim to provide a comprehensive
list of best practices to use in modeling reflection and subsequent spin
estimation to ensure high-quality and reliable constraints.

This paper is a step toward that goal. We do not attempt here to remeasure the
spins or to produce a definitive catalog. We emphasize that this article
is, by design, a synthesis of the workshop discussions rather than a
stand-alone science paper presenting new results: its goal is to define the
\emph{structure} of a quality-control standard---the pillars, the filtering
criteria, the tiered classification, and the reporting checklist---together
with a roadmap for its quantitative calibration. Instead, we propose a
framework for evaluating the quality of reflection-based spin measurements
reported in peer-reviewed publications, which at the same time can be used as a
list of best practices to inform future analyses. The framework is intended to
be strict enough to identify measurements that should not be used in a
high-confidence compilation, but flexible enough to acknowledge that different
types of sources and observing modes can impose different obstacles. Our
emphasis is on traceability: a reader should be able to understand why a
measurement is considered robust, why it is red-flagged, or why it cannot be
assessed from the published information. Although we recognize that alternative
criteria for reliability could be put forward, through our experience working
on many such analyses, we suggest the following standardization scheme for
assessing the quality of spin measurements while remaining open to adjustments
and redefinitions over time. The calibration of explicit numerical
thresholds for each criterion, which requires an extensive program of
simulations over realistic scenarios
(Section~\ref{demonstration-studies-needed}), will be presented in a dedicated
companion publication.

\section{Scope of the Problem}\label{scope}

Claims of ``spin measurements'' are widely found in the literature but often
with mixed levels of precision and/or accuracy. A spectral fit can return a
value of the dimensionless spin parameter, but the reliability of that number
depends on the data quality, the observability of the reflection signatures,
the ability to isolate the relativistic component, the assumptions made
{\it a priori} about physical parameters of the source (e.g., the disk
inclination, elemental abundances, or the coronal geometry), and the degree to
which alternative models have been explored. We therefore distinguish between a
\emph{reported spin value} and a \emph{reliable reflection-based spin
constraint}. The former is any published value obtained from a reflection
model; the latter is a value whose uncertainty budget is credible enough for
use in subsequent studies, such as population studies or in comparison with
other spin determinations.

The framework we propose herein follows four principles. First, quality
assessment should be applied to each specific measurement, not just to a
source. The same black hole binary can pass through hard, intermediate, and
soft states; the same active galactic nucleus (AGN) can be observed in epochs
with different absorption, flux levels, or reflection strengths. A given source
may therefore have a mixture of robust and unreliable spin estimates in the
literature.

Second, quality assessment should be based on published evidence. If a paper
does not provide enough information to determine whether a quality criterion is
satisfied, the measurement should not automatically be rejected as wrong;
instead, it should be flagged and labeled as \emph{non-assessable}. This
distinction is important. A not-assessable measurement may be accurate, but it
cannot be used in a transparent high-confidence compilation without additional
analysis.

Third, the primary classification should be conservative and discrete. A
continuous quality score is attractive, but it risks creating a false sense of
precision unless the weights are calibrated. We recommend a binary gate for
major failure modes, followed by a tiered quality label for measurements that
pass the gates. This allows the community to converge on a robust census while
still retaining information about additional measurements that are useful but
less secure.

Fourth, the framework should judge the published evidence and the
underlying assumptions, never the authors or research groups behind a
measurement. A measurement obtained from limited data---whether limited
in signal-to-noise, passband, or spectral resolution---may have been the best
possible analysis at the time. The purpose of the framework is not to
stigmatize, but to make clear which measurements are appropriate for which
downstream use. The quality labels in the curated compilation
(Section~\ref{toward-a-curated-spin-compilation}) must be traceable to
published evidence and must be assigned identically regardless of the
publication that reported the measurement.

\section{Pillars for a reliable reflection spin
constraint}\label{three-pillars-of-a-trustworthy-reflection-spin-constraint}

\subsection{Detectability}\label{detectability}

A reflection spin measurement first requires the relativistic reflection to be
significantly detected. Detectability combines two elements: the intrinsic
strength of the reflection component and the statistical quality of the
observation. A spectrum can have excellent signal-to-noise but negligible
inner-disk reflection, or strong reflection signatures but too few counts to
constrain the reflection's shape. Both cases are inadequate for accurate and
precise spin inference.

In practice, detectability should be assessed using evidence that is as
model-independent as possible. Useful diagnostics include residuals after
fitting a physically plausible continuum, the statistical improvement produced
by adding a relativistic reflection component (e.g., quantified by the change
in goodness of the fit), the equivalent width or flux of the broad Fe K
feature, the number of source counts in the Fe K band, the detection
significance of the Compton hump or smeared edge, and posterior evidence that
the relativistic component is required. 

One of the parameters in the reflection model, the reflection fraction---which
quantifies the fraction of photons reflected off the disk vs. those reaching
the observer directly---can be useful as a descriptive parameter, but it should
not by itself be used as the primary quality metric. This is because it is
model-dependent and can be affected by the assumed geometry, continuum, and
definition of the reflected flux adopted
\citep{Dauser2014,Dauser2016,Steiner2017}.

A measurement should be red-flagged if a smooth continuum curvature is
plausibly sufficient to describe the spectrum, rather than exhibiting
identifiable reflection features. This is especially important for claims of
very high-spin, where broadening can move part of the line flux to a smooth red
wing that may be confused with continuum structure, absorption curvature, or an
inadequately modeled thermal component at soft X-ray energies
\citep[see][]{Bambi2021}.

\subsection{Uniqueness, or separability}\label{uniqueness-or-separability}

Simply detecting reflection is not sufficient; the relativistic component must
be separable from other spectral components. The relevant question is: can the
data distinguish inner-disk reflection from distant reflection, absorption,
continuum curvature, instrumental features, and any other emission components?
We emphasize that separating a relativistic component from the other
model constituents is necessary but not sufficient, since the same spectral
features should not be described by an alternative viable model (for instance,
one dominated by complex absorption; see Section~\ref{degeneracies}). Thus,
uniqueness, in the sense intended here, does not claim that the reflection
solution is proven correct; only that competing descriptions are excluded at
the level the data allow.

Separability strongly depends on the passband and spectral resolution. Coverage
of the Fe K band constrains the line and edge structure; hard X-ray coverage
($>10$\,keV) constrains the Compton hump and the primary continuum; soft
coverage is essential when absorption (either cold and neutral or warm and
ionized), disk emission, or warm-corona components affect the lower-energy
continuum. {\it NuSTAR} transformed the field by providing focusing sensitivity
up to 79\,keV, allowing the Compton hump and high-energy continuum to be
measured directly from many sources \cite{Harrison2013}. However, {\it NuSTAR}
alone has limited spectral resolution in the Fe K band, so unresolved narrow
emission or absorption can still bias the inferred relativistic profile.
Conversely, instruments with excellent soft-band sensitivity but no hard X-ray
coverage can mistake continuum curvature for broad reflection if the Compton
hump is not constrained.

For AGN, the uniqueness of reflection signal modeling is often limited by the
presence of warm absorption, distant reflection, and narrow Fe K emission
produced by transmission through a relatively large column density (e.g.;
through the torus in Seyfert 2 AGN). For X-ray binaries, uniqueness is often
limited by rapid variability, disk winds, strong thermal emission, and commonly
detector pile-up given the brightness of Galactic X-ray binaries. In both
classes, the combination of high spectral resolution and broad passband is
ideal. {\it XRISM}/Resolve and future microcalorimeters will be particularly
valuable to distinguishing narrow absorption and emission components that can
masquerade as or distort the relativistic reflection signal
\citep{Brenneman2025,XRISM2024NGC4151,Draghis2025CygX1}.

\subsection{Robustness}\label{robustness}

Even when reflection is detected and isolated, spin inference can be
model-dependent. Robustness asks whether the spin remains stable under
plausible changes in the assumptions used to model the data. These assumptions
include the emissivity profile of the accretion disk and/or coronal geometry
\citep{Dauser2013,Nekrasov2025}, disk density \citep{Garcia2016,Tomsick2018},
ionization structure \citep{Svoboda2012}, iron abundance \citep{Garcia2018},
disk inclination \citep{Connors2019J1550}, disk thickness
\citep{TaylorReynolds2018}, returning radiation
\citep{Connors2020J1550,Dauser2022}, high-energy cutoff or coronal temperature
\citep{GarciaEcut2015,Kammoun2024HEXP}, and treatment of distant reflection and
absorption \citep{BarretCappi2019,Parker2022}. The distinction between
statistical precision and physical accuracy is central here. A fit can quote a
very small statistical uncertainty while ignoring systematic alternatives that
would significantly alter the inferred spin.

Robust analyses should therefore explore multiple plausible models and/or model
flavors, or justify why a restricted model family is adequate. For example, a
model that adopts a lamppost corona (e.g., \texttt{relxilllp}) may be
appropriate for some geometries, but an extended or equatorial corona can
produce different illumination patterns. A single-density parameterization of
the accretion disk gas may be inadequate for sources whose spectra require
higher disk densities or ionization gradients \citep{Garcia2016}. Thus, a
measurement that is stable across these reasonable choices is more reliable
than one that depends on a single untested model configuration.

Robustness also includes consistency across epochs and accretion states, but
this must be interpreted carefully. A black hole spin should be constant on
observational timescales, whereas the inferred inner disk radius may change if
the disk truncates or if the reflection model is no longer applicable. This is
the case in stellar-mass black hole binaries, where the disk truncation changes
over time during the hard state \citep[e.g.;][]{Garcia2015}, or where
reflection can be produced by returning disk radiation rather than the X-ray
corona, in which case standard reflection models are not applicable
\citep{Connors2020J1550}. Apparent spin changes across epochs are therefore
diagnostic of modeling or state-dependent systematics, not of physical spin
evolution.

\subsection{\texorpdfstring{Class-specific implementation of the
criteria}{Class-specific implementation of the
criteria}}\label{class-specific}

The three pillars are deliberately source-class-agnostic: detectability,
uniqueness, and robustness express requirements that any reflection-based spin
measurement must satisfy, and a shared framework is what enables a single
curated compilation spanning stellar-mass and supermassive black holes,
including comparisons with the gravitational-wave spin distribution. However,
the dominant systematic uncertainties differ substantially between the two
source classes, and the \emph{implementation} of each criterion must reflect
this. The quality tier of a measurement
(Section~\ref{recommended-classification-scheme}) is therefore always assigned
against the criteria appropriate to its source class.

For AGN, uniqueness is most commonly limited by warm and ionized absorption, by
distant reflection and narrow Fe~K emission produced in large-scale structures
such as the torus, and by the modeling of the soft excess; robustness is most
commonly limited by the treatment of these components, by the Eddington-ratio
regime, and by the combination of non-simultaneous epochs across long
variability timescales. The corresponding class-specific checks are: an
explicit absorption model, with an absorption-dominated alternative (e.g.,
partial covering) tested or excluded where warm absorption is known to be
present; a distant-reflection component whenever narrow Fe~K emission is
detected; and a demonstration that the adopted soft-excess treatment does not
drive the inferred spin.

For X-ray binaries, detectability and instrumental systematics are most
commonly limited by pile-up and instrument-mode effects, given the brightness
of Galactic sources; uniqueness by the thermal disk continuum and its interplay
with reflection, by disk winds, and by rapid variability; and robustness by
accretion-state dependence, including disk truncation in the hard state and
returning disk radiation in the soft and intermediate states
\citep{Connors2020J1550,Dauser2022}. The corresponding class-specific checks
are: a quantified pile-up budget with extraction-region tests; a truncation
test in the hard state, or an explicit lower-limit interpretation of the spin
(Sec.~\ref{accretion-state-is-incompatible-with-modeling-assumptions}); time-
or flux-resolved spectroscopy when strong variability is present; and a
demonstration that the reflection signal is separable from the thermal
continuum in disk-dominated states.

The forthcoming calibration campaign
(Section~\ref{demonstration-studies-needed}) will accordingly derive
class-specific thresholds, since a single set of numbers cannot be expected to
serve both regimes.

\section{Quality-control criteria}\label{quality-control-criteria}

We propose the following criteria as binary filters to identify problematic
measurements. A measurement that fails any one of these criteria should not be
included in a high-confidence spin compilation unless the study provides a
convincing mitigation. Measurements that cannot be evaluated because the
required information is missing should be flagged as not assessable rather than
fully rejected.

\subsection{Reflection is not significantly
detected}\label{reflection-is-not-significantly-detected}

A spin estimate should be red-flagged if the paper does not demonstrate a
significant relativistic reflection component. Evidence for the presence of
reflection may include residuals to a continuum-only model, a significant
improvement in fit statistic when relativistic reflection is added, a broad Fe
K line or smeared edge that cannot be explained by narrow reflection or
absorption, and a hard-band excess consistent with a Compton hump. The relevant
statistic may be $\Delta\chi^2$, $\Delta C$-stat, Bayesian evidence, posterior
predictive checks, or simulation-calibrated significance. Because different
studies report different quantities, the assessment should record which
evidence was actually available.

A practical implementation is to require at least one of the following: (i) a
clearly shown broad Fe K residual after continuum removal; (ii) a statistically
significant fit improvement from adding relativistic reflection; (iii) a
well-constrained broad Fe K equivalent width or line flux above the continuum;
or (iv) a posterior in which the relativistic reflection normalization is
inconsistent with zero and the spin constraint is not pegged solely by a
boundary condition.

\subsection{Limited energy coverage}\label{limited-energy-coverage}

A reliable reflection spin measurement generally requires constraints on both
the Fe K band and the hard continuum/Compton hump. A measurement should be
red-flagged if it relies only on a narrow energy range that cannot separate the
continuum, the relativistic line, the edge, and the Compton hump. As a
practical rule, robust measurements should include either hard X-ray coverage
extending to at least the 25--30\,keV range with adequate source counts
above background---as a practical guideline, the source should remain
detected above the background up to those energies with enough counts to permit
reliable fit statistics; for $\chi^2$ fitting this corresponds to $\gtrsim 20$
background-subtracted counts per spectral bin without requiring a rebinning so
coarse that the Fe~K or Compton-hump structure is washed out, whereas
Poisson-based statistics (e.g., the Cash statistic) can operate reliably at of
order one source count per channel---, unless the source is in a particular
configuration such that the observation demonstrably constrains the continuum
and reflection components. This could be the case of sources with a very cold
X-ray corona, in which case the high-energy cutoff in the power-law continuum
is observed at energies close or below the Compton hump, which effectively
reduces its intensity, due to the scarcity of high-energy photons
\citep[e.g.;][]{Kara2017}.

This criterion should not be applied mechanically. For some high-redshift AGN
the rest-frame Compton hump can shift into a lower observed energy band,
although the limited effective area of current X-ray observatories does not
allow for a large number of such cases. In contrast, hard-band coverage alone
is not always enough. A fit based only on {\it NuSTAR} data can constrain the
Compton hump and continuum but may lack the Fe K resolution needed to separate
broad and narrow components. At the same time, fits using only hard X-ray
data can be adequate in specific cases, such as in heavily absorbed sources
whose soft band is strongly suppressed by a large column density. The
preferred configuration is simultaneous or strictly contemporaneous
soft-plus-hard coverage, with the importance of soft coverage increasing in
sources with warm absorption, disk winds, or strong thermal components.

\subsection{Observational data are dominated by known
systematics}\label{observational-data-are-dominated-by-known-systematics}

A measurement should be red-flagged if the data used for the spin inference are
dominated by detector or calibration effects that can distort the relevant
spectral features. Examples include severe pile-up, inadequately treated
timing-mode calibration, hard-band data dominated by background while claiming
a Compton-hump constraint, unexplained slope differences among instruments,
gain shifts that modify the Fe K region, and instrumental features that require
systematic uncertainties larger than the claimed spin precision.

Pile-up is especially important for bright X-ray binaries observed with
charged-couple devices (CCDs)
\citep[e.g.;][]{Miller2006GX339,DoneDiazTrigo2010}. Excluding the core of the
point-spread function (PSF) can mitigate pile-up, but may introduce calibration
uncertainties in the wings \emph{(N. Schartel, private comm.)}. The exact
threshold for unacceptable pile-up depends on the instrument and mode, but
analyses should explicitly quantify the pile-up fraction, the mitigation
strategy, and the effect of alternative extraction regions on the inferred
reflection profile. If this information is absent, the measurement should be
treated as not assessable.

The simultaneity of multi-instrument data is also part of this criterion.
Non-simultaneous soft and hard observations can be dangerous for variable
sources, especially in X-ray binaries and rapidly variable AGN. A broadband fit
constructed from observations separated by many source variability timescales
can create artificial curvature or erase the real spectral structure. When
non-simultaneous data are used, the analysis should demonstrate spectral
stability or model the variability explicitly and adequately.

\subsection{Accretion state is incompatible with modeling
assumptions}\label{accretion-state-is-incompatible-with-modeling-assumptions}

Reflection-based spin inference assumes that the observed relativistic
reflection originates from an optically thick disk whose inner radius is
physically tied to the ISCO. This assumption must be examined in light of the
source accretion state.

In the hard state of X-ray binaries, the disk may be truncated outside the ISCO
\citep[e.g.;][]{BasakZdziarski2016,Dzielak2019}, although the same source and
even the same data have also been modeled and interpreted as reaching the ISCO
\citep[e.g.][]{Garcia2015,Fuerst2015,WangJi2018}. A hard-state spin measurement
is therefore only reliable if the analysis demonstrates that the reflecting
disk reaches the ISCO or treats the inner radius separately from spin. If the
model assumes the disk reaches the ISCO without testing truncation, the
resulting spin should be red-flagged and at the very least considered as a
lower limit, depending on the particular assumptions made during the fitting
procedure\footnote{For example, if the spin is measured in a truncated disk
assuming the inner radius is at the ISCO, the model will likely artificially
lower the spin to compensate for the improper assumption, in which case the
spin determination can be regarded as a lower limit at best.}.

In intermediate states, rapid spectral variability can invalidate a
time-averaged spectrum if the continuum, reflection fraction, ionization, or
inner radius changes during observation exposure
\citep[e.g.;][]{Sridhar2020,AxelssonVeledina2021}. Time-resolved or
flux-resolved spectroscopy may be required. A measurement should be red-flagged
if strong variability is present but is ignored or not properly taken
into account.

In soft states, thermal disk emission can dominate the spectrum, and thus the
reflection signal may be weak or entangled with disk spectral complexity. Most
standard reflection models like \relxill\ do not always include the full
physics of a strongly illuminated, thermally emitting disk---although other
models do, such as a version of the \textsc{reflionx} model, called
\textsc{refbhb} \citep{RossFabian2007}. Soft-state reflection spin estimates
should therefore be considered robust only if the analysis demonstrates that
the reflection signatures are detectable and separable from the thermal
continuum and that the adopted model accounts for the relevant disk emission
physics.

AGN do not seem to map one-to-one onto binary accretion states, but analogous
concerns apply: high Eddington ratio sources, strong soft excesses, variable
absorbers, and complex coronae can all undermine simple reflection
interpretations if not modeled carefully.

\subsection{Modeling choices are too restrictive or outdated for the data
quality}\label{modeling-choices-are-too-restrictive-or-outdated-for-the-data-quality}

A measurement should be red-flagged if the model used is known to be inadequate
for the data quality or scientific claim. This includes fits that use a
phenomenological line profile when a self-consistent reflection continuum is
required, fits that ignore distant reflection while claiming a broad-line spin,
fits that adopt outdated reflection grids without checking whether updated
physics changes the result, and fits that quote high precision while fixing
parameters that are known to be degenerate with spin (e.g., inner radius,
inclination, emissivity index, and/or coronal height).

At the same time, older measurements should not be dismissed solely because
they used older models. Many remain valuable. However, the quality label should
distinguish between measurements that have been confirmed with modern models
and broadband data, and measurements whose systematic uncertainty is unknown
because the analysis predates important model developments. A good example is
the famous AGN MCG$-$6-30-15, which has been reported to have a near-maximal
spin measured using a variety of models and datasets since 2006, both with and
without the benefit of {\it NuSTAR} (extended passband) and {\it XRISM} (<5\,eV
spectral resolution in the Fe K band), and with the implementation of the first
models produced for reflection
\citep[e.g.,][]{Brenneman2006,Chiang2011,Marinucci2014}. This demonstrated that
if a source exhibits a Fe K emission profile strong enough and sufficiently
broad enough (coupled with good enough data to adequately model the warm
absorber), spin measurements are relatively easy to obtain. But certainly the
accuracy of the measurement is greater once one has access to the best data
across the entire passband.

\subsection{Statistical reporting is incomplete or
misleading}\label{statistical-reporting-is-incomplete-or-misleading}

A measurement should not be treated as high-confidence unless the uncertainty
reporting is commensurate with the complexity of the problem. Spin
constraints that are quoted with extreme precision should be flagged and looked
in more detail. Often spin uncertainties are quoted only reflecting the
statistical component, which arises from data noise and model priors. This
mainly shows the precision of the measurement. However, systematic effects such
as biases introduced by incorrect model assumptions, instead impact accuracy,
and are rarely folded into the reported constraints. High-confidence
measurements should therefore state explicitly whether the quoted uncertainty
is purely statistical and, where possible, estimate the systematic
contribution. Boundaries and one-sided limits must be clearly identified. If
the best fit is pegged at the maximum allowed spin, the result should be
reported as a lower limit unless the posterior is demonstrably bounded away
from the maximum by the data rather than by the model grid. Robust analyses
should provide confidence contours or posterior samples involving spin and key
covariant parameters such as disk inclination, iron abundance, ionization,
emissivity, density, reflection fraction, and inner radius.

\section{Recommended classification
scheme}\label{recommended-classification-scheme}

We recommend classifying each reported spin value at the level of a
source-observation-model combination. A single paper may therefore contribute
more than one measurement and a single source may contain measurements in
multiple quality levels.

The tiers are assigned through an explicit rule set operating on the
binary filters of Section~\ref{quality-control-criteria}. We designate two of
the six criteria as \emph{critical}: the detectability requirement
(Sec.~\ref{reflection-is-not-significantly-detected}) and the
instrumental-systematics requirement
(Sec.~\ref{observational-data-are-dominated-by-known-systematics}), since a
measurement failing either cannot be salvaged by strengths elsewhere. The
remaining criteria (passband, accretion state, model robustness, and
statistical reporting) are \emph{non-critical}: individual shortfalls can in
principle be mitigated or absorbed into an enlarged error budget.

\textbf{Tier A: high-confidence measurement.} The measurement passes all
six criteria of Section~\ref{quality-control-criteria}, and the robustness
tests of the reporting checklist (items 7--8 in
Section~\ref{recommended-minimum-reporting-checklist-for-future-reflection-spin-analyses})
are reported together with their outcomes: the spin is shown to be stable under
plausible model variants and the relevant parameter covariances are provided.
Tier A measurements are appropriate for population studies and mission-level
forecasts.

\textbf{Tier B: usable but systematics-limited measurement.} The
measurement passes both critical criteria and fails at most two non-critical
criteria, each with a documented mitigation or an explicit enlargement of the
systematic error budget (e.g., incomplete passband, insufficient spectral
resolution in the Fe~K band, limited model exploration, moderate unresolved
absorption, or imperfect simultaneity between soft and hard coverage).
Tier B measurements can be used in population work only if their enlarged
systematic uncertainties are correctly propagated.

\textbf{Tier C: provisional or red-flagged measurement.} The measurement
fails at least one critical criterion, fails more than two non-critical
criteria without mitigation, or the reported spin depends on an assumption that
the data do not test (e.g., an ISCO-reaching disk assumed in the hard state
without a truncation test;
Sec.~\ref{accretion-state-is-incompatible-with-modeling-assumptions}). Tier C
measurements should not be used in high-confidence compilations, although they
may be useful as case studies or for planning improved observations.

\textbf{Tier U: not assessable from the publication.} The study does not
provide the information required to evaluate at least one critical criterion.
Tier U reflects traceability rather than reliability: it flags measurements
that cannot be evaluated from the publication alone, but may nonetheless be
accurate.

We note that a residual element of expert judgment is irreducible---most
notably, deciding whether a documented mitigation is convincing---which is
precisely why we recommend an editorial review model for the curated
compilation (Section~\ref{toward-a-curated-spin-compilation}) rather than a
fully automated classification. Full numerical objectivity, in which each
criterion is evaluated against calibrated thresholds, awaits the simulation
program of Section~\ref{demonstration-studies-needed} and will be presented in
the companion publication.

\begin{table}[htbp]
\caption{Proposed quality matrix for each reported reflection-based spin
measurement.}\label{tab:quality-matrix}
\small
\begin{tabular}{p{0.16\textwidth}p{0.30\textwidth}p{0.32\textwidth}p{0.13\textwidth}}
\hline
\textbf{Category} & \textbf{Minimum evidence} & \textbf{Failure mode} & \textbf{Recommended label} \\
\hline
Detectability & Broad Fe K residual, significant fit improvement, or reflection posterior inconsistent with zero & Spin driven by continuum curvature or unconstrained reflection normalization & C or U \\
Passband & Fe K band plus hard continuum/Compton hump, preferably simultaneous soft+hard coverage & Fe-line-only or hard-band-only fit; hard-band background dominated; no continuum anchor; unaccounted variability between non-simultaneous soft and hard observations& B, C, or U \\
Instrumental systematics & Pile-up, background, calibration, gain, and cross-normalization assessed & Distorted line profile or artificial curvature (e.g., pile-up in bright XRBs) & C or U \\
Accretion state & State identified and model assumptions justified & Disk truncation, strong variability, or soft-state disk dominance ignored (XRBs); unmodeled absorption or soft-excess complexity (AGN) & B or C \\
Model robustness & Alternative geometries/physics tested or justified & Spin changes under plausible model variants & B or C \\
Reporting & Error bars, limits, pegging, fit statistics, and covariances reported & Over-precise or boundary-driven spin value & B, C, or U \\
\hline
\end{tabular}
\end{table}

\section{Recommended minimum reporting checklist for future reflection spin
analyses}\label{recommended-minimum-reporting-checklist-for-future-reflection-spin-analyses}

To make future measurements assessable, studies should report the following
information as a matter of standard practice.

\begin{enumerate}
\def\labelenumi{\arabic{enumi}.}
\tightlist
\item
  \textbf{Observation metadata:} observatory, instrument, mode, ObsID,
exposure, count rate, extraction regions, background treatment, and
simultaneity of multi-instrument data.
\item
  \textbf{State and variability diagnostics:} hardness ratio, flux state,
timing properties when relevant, light curves, and justification for
time-averaged versus time-resolved spectroscopy.
\item
  \textbf{Instrumental-systematic checks:} pile-up estimate, mitigation
strategy, gain shifts, calibration caveats, cross-normalization constants,
background dominance at high energies, and the effect of alternative extraction
choices.
\item
  \textbf{Continuum definition:} continuum model, energy ranges used to
establish it, residual plots, and sensitivity of the reflection result to
continuum alternatives.
\item
  \textbf{Reflection detection evidence:} $\Delta\chi^2$/$\Delta C$-stat or
Bayesian evidence for adding relativistic reflection, broad-line equivalent
width or flux, Compton-hump detection significance when applicable, and
residual plots before and after adding reflection.
\item
  \textbf{Component separability:} treatment of distant reflection, narrow Fe K
emission, warm absorption, disk winds, soft excess, thermal disk emission, and
other source-specific components.
\item
  \textbf{Model robustness tests:} at minimum, tests of emissivity
prescription, coronal geometry, disk density, iron abundance, ionization
treatment, inclination, cutoff energy, and inner radius assumptions when these
parameters are relevant.
\item
  \textbf{Covariances of the parameters:} confidence contours or posterior
samples for spin versus key degenerate parameters, and clear identification of
pegged parameters or one-sided limits.
\end{enumerate}

Other desired information that would be convenient to collect from future
studies has to do with the reproducibility of the results by providing the
required products and data files. This includes model files, scripts with the
model fitting procedure in the software of choice (e.g.; XSPEC, ISIS, Sherpa,
etc.), response files or links to public archives, posterior chains when
available, and machine-readable tables of best-fit parameters.

\section{Demonstration studies needed}\label{demonstration-studies-needed}

The above criteria can now be applied qualitatively, but quantitative
thresholds require calibration, and we contend that such thresholds must
be \emph{derived} rather than asserted. The relevant quantities (detection
significance, minimum counts, equivalent-width floors, passband requirements,
robustness tolerances) are not universal constants: as the demonstrations below
show, they depend on the spin value itself, on the coronal geometry, on the
source flux, and on the source class, among other system properties. Quoting a
single uncalibrated set of numbers would risk those values acquiring unearned
authority and being applied mechanically regardless of context. We
recommend---and are undertaking, for a dedicated companion
publication---a focused set of demonstration simulations spanning
realistic scenarios in source class, spin, inclination, coronal geometry, flux,
accretion state, absorption complexity, and instrument configuration, rather
than immediate claims of universal thresholds. These simulations should be
framed as examples of failure modes, not as definitive boundaries; the
demonstrations presented in this section illustrate the methodology of that
program.

A first simulation set should examine the passband. One can simulate spectra
with known spin, reflection strength, inclination, and coronal temperature,
then fit the same data with (i) only soft/Fe-band coverage, (ii) only {\it
NuSTAR}-like hard coverage, and (iii) simultaneous soft/hard coverage. The goal
is to show how the posterior on the spin changes when the Compton hump or the
Fe K structure is removed.

A second set should examine reflection detectability. By varying reflection
strength and exposure time, one can identify regimes where a formal spin value
is returned even though the broad line and/or Compton hump signal is weak. This
would calibrate approximate requirements in terms of counts in the Fe K
feature, equivalent width, or reflection component detection significance. We
also emphasize that our ability to constrain the spin at a desired precision is
a function of the spin value itself; higher spins are easier to measure given
their inherently broader Fe K emission \citep{Piotrowska2024HEXP}; low spin
solutions are naturally degenerated with the system geometry, among other
things.

\begin{figure}[H]
\centering
\includegraphics[width=0.8\textwidth]{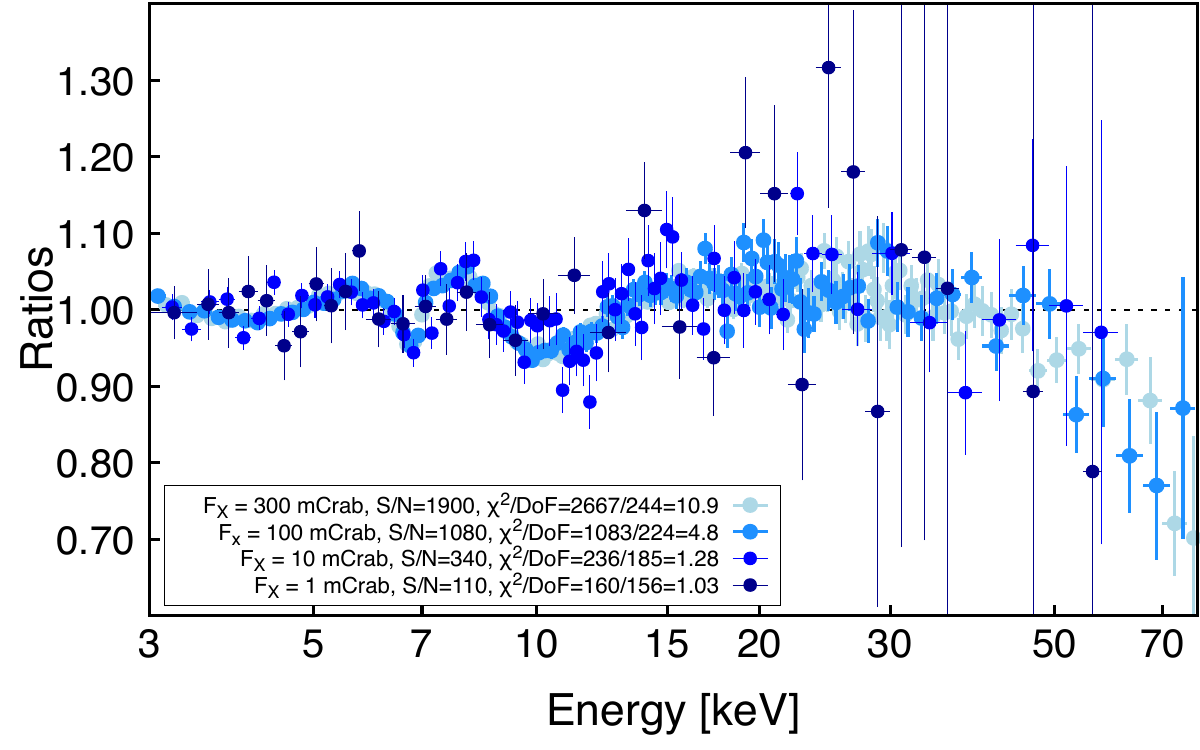}
\caption{Ratio spectra of fits of Schwarzschild solutions ($a_*=0$) to
simulated {\it NuSTAR} spectra based on maximal intrinsic spin ($a_*=0.998$).
The simulated spectra are all based upon an exposure time of 20 ks, but at
variable intrinsic source flux.\label{fig:sn_ratios}}
\end{figure}

As a demonstration of such tests, Figure~\ref{fig:sn_ratios} shows the
importance of intrinsic source counts for a detectable reflection spectrum.
Here we simulate {\it NuSTAR} spectra at different source fluxes whereby the
black hole spin is assumed to be maximal. We then fit the spectra with a
zero-spin model, i.e., a Schwarzschild black hole. Once the S/N falls to the
order of hundreds, one can reproduce the data well with a zero-spin model. Spin
constraints in the literature often derive from similar quality fits, and the
$S/N$ of the data is rarely a significant consideration of those constraints,
and yet Figure~\ref{fig:sn_ratios} is a simple demonstration of just how
important $S/N$ is.

\begin{figure}[H]
\centering
\includegraphics[width=0.8\textwidth]{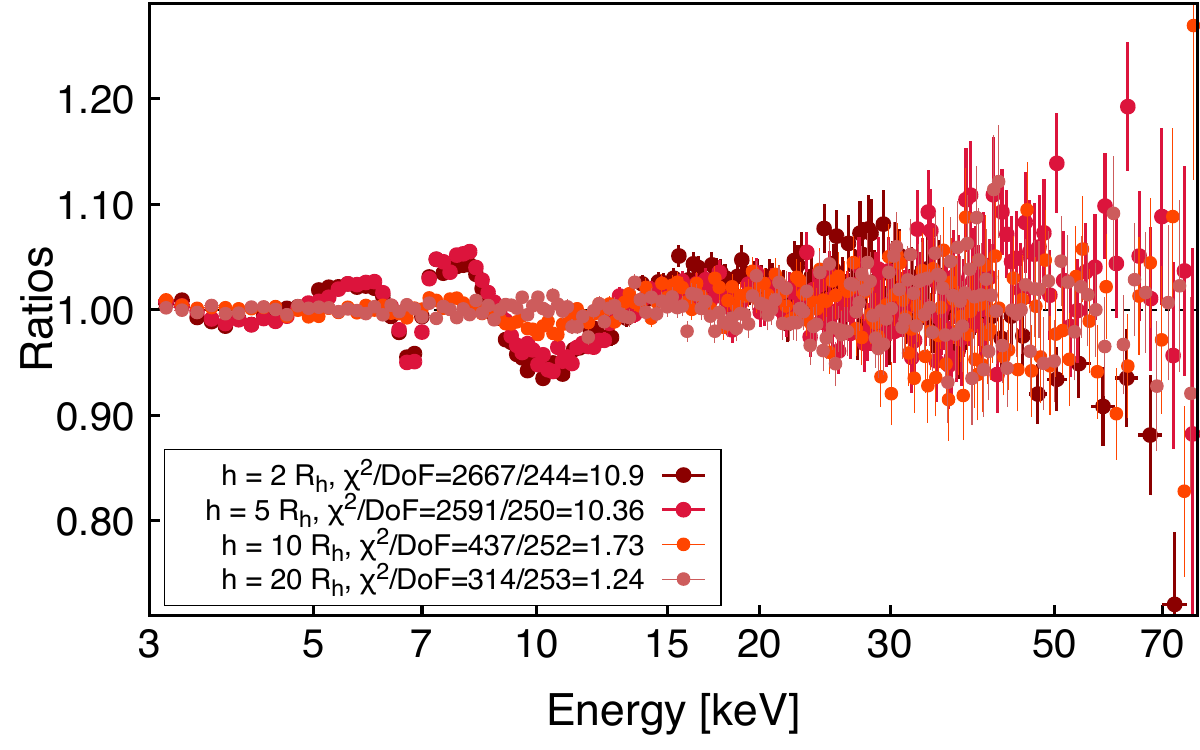}
\caption{Ratio spectra of fits of Schwarzschild solutions ($a_*=0$) to
simulated {\it NuSTAR} spectra based on maximal intrinsic spin ($a_*=0.998$).
The simulated spectra are all based upon an exposure time of 20\,ks, but at
variable intrinsic lamppost height (specified in terms of the horizon radius,
$R_h$), and by association, both reflection fraction and emissivity across the
accretion disk. The intrinsic ($2-10$\,keV) source flux is
300\,mCrab.\label{fig:h_ratios}}
\end{figure}

Figure~\ref{fig:h_ratios} shows a similar test, this time assuming variation in
the intrinsic coronal lamppost source height, which informs both the reflection
fraction and the emissivity across the disk. Again, here it is clear that
almost any high spin constraint can be highly suspect if one allows freedom in
the coronal geometry. The impact of the system geometry is rarely carefully
considered in the literature, and yet the source emissivity itself is a key
component with regards to the detectability of black hole spin.

A third set should examine component confusion. Simulated spectra should
include narrow distant reflection, warm absorption, disk winds, and variable
continuum curvature. The same spectra can then be fitted with incomplete models
to quantify biases. This is another situation in which the answer is also
likely spin-dependent. This is particularly true when separating distant
reflection from relativistic reflection produced near a lowly-spinning black
hole. In this case, the two components produce an inherently narrow Fe K
emission. Evidently, access to high-energy resolution data (e.g.; {\it XRISM})
can break (or at least minimize) this degeneracy.

A fourth set should examine model physics. Synthetic spectra generated from
high-density disks, non-lamppost illumination, ionization gradients, finite
disk thickness, or GRMHD-based disk structures should be fit with standard
models. Recent work using GRMHD-informed spectra has already shown that model
assumptions can affect the shape of the reflected spectrum
\citep[e.g.;][]{Kinch2016,Nagele2026}, and thus possibly the spin recovery,
especially when the simulated accretion geometry differs from the fitting model
\cite{Shashank2025}. These recovery tests will explicitly include the
iron-abundance/spin and disk-density/spin degeneracies discussed in
Section~\ref{degeneracies}.

The purpose of these simulations is not to completely invalidate existing
measurements. Rather, they provide a controlled basis for deciding which
missing checks are likely to matter for a given class of data.

\subsection{\texorpdfstring{Known parameter degeneracies and their place
in the framework}{Known parameter degeneracies and their place in the
framework}}\label{degeneracies}

The demonstrations above target signal-to-noise and coronal geometry, but
the literature has identified several additional degeneracies that any quality
framework must confront, and which the calibration program must quantify. We
briefly review the four most widely discussed, and identify for each the
criterion of Section~\ref{quality-control-criteria} and the reporting-checklist
item designed to catch it. The framework is intended to absorb these known
systematics, not to replace the detailed studies that characterized them.

\emph{Warm-absorber/reflection degeneracy.} In several key AGN, the broad
red wing attributed to relativistic reflection can alternatively be described
by complex, partially covering absorption, leading to long-running debates over
sources such as MCG$-$6-30-15
\citep{MillerTurnerReeves2008,MillerTurnerReeves2009,Reynolds2009,Marinucci2014}
and 1H\,0707$-$495 \citep{Fabian2009,Hagino2016}. High-resolution
microcalorimeter spectroscopy with {\it XRISM}/Resolve is now directly
resolving absorber structure in such sources
\citep{Brenneman2025,XRISM2024NGC4151}, sharply reducing this degeneracy.
Within our framework this is a failure of \emph{uniqueness}: the detectability
criterion (Sec.~\ref{reflection-is-not-significantly-detected}) requires that
an absorption- or continuum-only description be excluded, and checklist item~6
requires the absorption treatment to be reported. The third simulation set
above is designed to calibrate this failure mode.

\emph{Pile-up.} Photon pile-up in CCD observations of bright sources
distorts the Fe~K profile and the continuum shape, and has been shown to bias
relativistic-line parameters in either direction depending on the mitigation
adopted \citep{Miller2010pileup,DoneDiazTrigo2010}. This is an
\emph{instrumental systematics} failure caught by the criterion of
Sec.~\ref{observational-data-are-dominated-by-known-systematics} and checklist
item~3, which already require a quantified pile-up fraction, the mitigation
strategy, and the effect of alternative extraction regions.

\emph{Iron-abundance/spin degeneracy.} Reflection fits frequently return
strongly supersolar iron abundances that are physically difficult to motivate
and are coupled to the inferred reflection strength and spin
\citep{Garcia2018}. A spin measurement that changes significantly between
fixed-solar and free-abundance fits is not robust. This degeneracy maps onto
the \emph{robustness} pillar (Sec.~\ref{robustness}) and checklist item~7,
which requires iron-abundance tests among the minimum model robustness checks.

\emph{Disk-density/spin degeneracy.} The historical assumption of a fixed
disk density ($n_e = 10^{15}$\,cm$^{-3}$) alters the low-energy reflection
continuum and the ionization balance, biasing the inferred iron abundance and
the parameters that covary with spin; high-density models partially resolve the
supersolar-abundance problem in both AGN and X-ray binaries
\citep{Garcia2016,Tomsick2018,Jiang2019AGN,Jiang2019XRB}. This likewise maps
onto checklist item~7 (disk density among the required robustness tests), and
it is an explicit target of the fourth simulation set above: recovery tests in
which high-density or free-abundance input spectra are fitted with
fixed-density, fixed-abundance models will quantify the resulting spin bias as
a function of data quality in the companion publication.

\section{Toward a curated spin compilation}\label{toward-a-curated-spin-compilation}

A high-value outcome of this effort would be a curated table of
reflection-based spin measurements with explicit quality labels. Such a
compilation should be built from a reproducible literature search, but it
should not rely exclusively on automated extraction. Automated tools can
identify candidate papers, sources, ObsIDs, reported spin values, error bars,
model names, and obvious boundary conditions. Expert review is still needed to
evaluate diagnostic figures, observation modes, state classifications, and
modeling assumptions.

The appropriate unit of this scheme is the \emph{measurement}, not the paper.
Required columns should include source name, source class, black hole mass
range, observation date, ObsID, instruments, energy band, simultaneity flag,
accretion state, model family, reported spin, uncertainty type, boundary flag,
evidence for reflection detection, passband flag, instrumental-systematics
flag, accretion state flag, robustness tests performed, quality tier, and
notes.

The compilation should also retain measurements that fail or cannot be
assessed. Excluding them silently would erase useful information about the
state of the literature. Instead, the table should allow users to filter for
Tier A measurements while still seeing why other reported values are
provisional.

A living repository is attractive because new measurements, reanalyses, and
calibration updates will continue to appear. However, a living product must
preserve credibility. We recommend an editorial model rather than an
unrestricted community Wiki style: submissions can be proposed by the
community, but changes to quality labels should be reviewed by a small rotating
board with expertise in reflection spectroscopy, instrumentation, and accretion
physics. Each version of the {\em spins list} should be citable and archived,
with a change-log documenting why measurements are added, removed, or
reclassified.

\section{Implications for future
missions}\label{implications-for-future-missions}

The quality-control framework has immediate implications for mission planning.
The Wake Forest workshop highlighted the need to propagate theoretical and
observational uncertainties into mission-level requirements. For reflection
spectroscopy, this means that future missions should not be evaluated only by
their ability to collect many photons. They must also provide the combination
of passband, spectral resolution, timing capability, calibration stability, and
observing flexibility needed to satisfy the detectability, uniqueness, and
robustness criteria. Evidently, this combination of the desired observing
capabilities represents an ideal scenario. In reality, financial and
programmatic considerations will certainly necessitate a continued need to rely
on a combination of data obtained from multiple telescopes and instruments,
likely through simultaneous observing campaigns.

The high spectral resolution in the Fe K band will improve separability by
resolving narrow absorption and emission features. This is already true for
studies featuring observations with {\it XRISM}
\citep{Brenneman2025,XRISM2024NGC4151,Draghis2025CygX1}. Broad hard-X-ray
coverage will remain essential for constraining the Compton hump and the
illuminating continuum. Simultaneous or coordinated soft-plus-hard observations
will be critical for variable sources. Large collecting area will allow
time-resolved spectroscopy, reducing biases from averaging over rapidly
changing states. Finally, mission simulations should include realistic source
complexity: absorption, winds, distant reflectors, extended coronae, ionization
gradients, disk density structure, and variability
\citep[e.g.][]{Garcia2024HEXP,Connors2024HEXP,Piotrowska2024HEXP,Kammoun2024HEXP,Ludlam2024HEXP}.

These requirements are especially relevant for {\it XRISM}, {\it NewAthena},
broadband concepts similar to {\it HEX-P}, and future high-throughput
microcalorimeter missions. The strongest spin constraints will come from not
any single capability, but from combinations that make the reflection signal
detectable, separable, and robust.

\section{Conclusions}\label{conclusions}

Reflection spectroscopy remains a powerful method for measuring black hole
spin, but there is an imperative need for the field to provide a more explicit
standard for deciding when a reported spin value is reliable. We propose a
practical framework organized around detectability, uniqueness, and robustness.
These principles lead to practical filters involving reflection significance,
passband, instrumental systematics, accretion state compatibility, model
exploration, and statistical reporting.

The immediate recommendation is to use these criteria to classify published
measurements into high-confidence, usable but yet systematics-limited,
provisional/red-flagged, and not-assessable tiers. The longer-term
recommendation is to build a curated, versioned, and community-maintained
compilation of reflection spin measurements, with each quality label traceable
to published evidence. Such a product would support population studies,
comparisons with gravitational-wave spin distributions, and mission planning
for the next generation of X-ray observatories.

A reliable spin measurement requires more than a small statistical error bar: a
detectable relativistic reflection signal, a demonstrable separation from
competing spectral components, and stability against plausible observational
and theoretical systematics.




\dataavailability{No new observational data were generated in this manuscript.
The proposed future spin-compilation table and assessment products should be
made available in a versioned public repository when produced.}

\acknowledgments{This draft is based on discussions motivated by the Wake
Forest workshop \emph{Recent Progress on Black Hole Spin Measurements Across
the Electromagnetic and Gravitational Spectra}. We thank the workshop
organizers, invited speakers, and participants for discussions that emphasized
the need for transparent, cross-method standards in black-hole spin inference.
}

\conflictsofinterest{The authors declare no conflicts of interest.}

\reftitle{References}

\bibliography{references}

\end{document}